\documentclass[sn-mathphys]{sn-jnl}

\jyear{2022}%

\theoremstyle{thmstyleone}%
\theoremstyle{thmstyletwo}%

\theoremstyle{thmstylethree}%

\begin{document}

\title[The Shadow of Charged Traversable Wormholes]{The Shadow of Charged Traversable Wormholes}


\author*[1,2]{\fnm{Mário} \sur{Raia Neto}}\email{mraianeto@estudante.ufscar.br}

\author*[3]{\fnm{Daniela} \sur{Pérez}}\email{daniela.perez2812@gmail.com}

\author*[4]{\fnm{Joaquín} \sur{Pelle}}\email{jpelle@mi.unc.edu.ar
}

\affil[1]{\orgdiv{Departamento de Física}, \orgname{Universidade Federal de São Carlos}, \orgaddress{\state{São Paulo}, \country{Brazil}}}

\affil[2]{\orgname{Instituto Nacional de Pesquisas Espaciais}, \orgaddress{ \state{São Paulo}, \country{Brazil}}}

\affil[3]{\orgdiv{Instituto Argentino de Radioastronom\'ia},\orgname{(IAR, CONICET/CIC/UNLP)}, \orgaddress{\state{Buenos Aires}, \country{Argentina}}}

\affil[4]{\orgdiv{Facultad de Matem\'atica, Astronom\'ia, F\'isica y Computaci\'on,}, \orgname{Universidad Nacional de C\'ordoba}, \orgaddress{\country{Argentina}}}


\abstract{We compute the shadow cast by a charged Morris-Thorne wormhole when the light source is a star located beyond  the mouth which is opposite to the observer. First, we provide an extensive analysis of the geodesic properties of the spacetime, both for null and massive particles. The geometrical properties of this solution are such that independently of the viewing angle, some light rays always reach the observer. Additionally, the structure of the images is preserved among the different values of the charge and scales proportionally to the charge value.
}





\maketitle

\section{Introduction}

Wormholes are exact solutions of Einstein Field Equations (EFE), which
connects two spacetime regions by a throat. The mouths are not hidden by event horizons, as in the case of black holes, and there are no singularities.



The first work on what we now call wormhole was published by A. Einstein and N. Rosen \cite{erb} in 1935. They tried to construct a model for an elementary particle that could be everywhere finite and free of singularities. Though their particle model was a complete failure, they arrived to the first wormhole spacetime model that is known as Einstein-Rosen bridge.

Almost twenty years passed until the subject was revived by J.A.Wheeler \cite{wheeler55} in 1955 when he published a work about ``geons''. This is the first time that a wormhole space diagram appears in the scientific literature. Two years later, C.W. Misner and J.A.Wheeler \cite{wheeler+57} analyzed the geometry of manifolds with non-trivial topology with the aim to explain completely all classical physics. The word ``wormhole'' is employed for the first time in the context of General Relativity.


In the seventies, H.G. Ellis introduced\cite{ellis1, ellis2} a kind of wormhole solution, called Ellis drainhole;  Bronnikov\cite{bronn} noticed that the solution was free from event horizons and singularities, geodesically complete and able to be ``crossed" independent of direction\footnote{There are two different solutions of the Ellis drainhole, with matter \cite{ellis1} and no matter flow \cite{ellis2}, being the later the simplest. Notice that both Bronnikov and Ellis arrived  independently to these solutions, so these are referred as ``Ellis-Bronnikov Wormholes".}. During the sixties and seventies, there was great progress in different aspects of General Relativity. However, not much work was focused on the analysis of Lorentzian wormholes. 

The pioneer work of Michael Morris and Kip Thorne \cite{mt} on \emph{traversable wormholes}, also called Morris-Thorne Wormholes (MTWH), caused a major revival on the subject. In their famous paper published in 1988, they focused on the necessary conditions in order to have a wormhole geometry that could connect two flat asymptotically spacetime regions; these solutions must not contain horizons or real singularities, and are connected by a ``throat'' that is kept open due to presence of exotic matter (matter that violates the energy conditions) that exerts gravitational repulsion\footnote{In 1989, G. Clément \cite{cle1} demonstrated that the Ellis drainhole is type of a traversable wormhole.}.


Much has been discovered about these astonishing solutions. Roman, proposed a mechanism to ``inflate" the microscopic wormholes discussed by Wheeler\cite{roman}; Kim has found cosmological traversable wormholes\cite{kim2}, Teo discovered the rotational version of MTWH\cite{teo} and A. Alias proposed a slowly rotational version of Teo's rotating metric\cite{alias}. Furthermore, Kim also constructed a form of charged MTWH\cite{kim}. A scalar field supporting the traversable wormhole was introduced by L.Butcher\cite{but} and a few years ago Lobo, Quinet and Oliveira  discovered the deSitter MTWH\cite{quinet}. Traversable wormholes have also been studied in alternative theories of gravitation such as $f(R)$ and $f(R,T)$ gravity\cite{cat,sant,ann,sam,dixit,god,mish,gau,bar,sharma,moraes,vip,mandal} and in analogue gravity\cite{eu}. There are works on traversable wormholes in cosmology\cite{tang,rah}, astrophysics \cite{dai}, thermodynamics\cite{rom,hong,termworm} and their shadows were also computed \cite{galin}.

Though several wormhole solutions have been extensively analyzed during the years, the first charged transversable wormhole geometry found more than twenty years ago\cite{kim} has barely been investigated. One interesting feature of this solution is that the presence of charge allows to minimize the content of exotic matter at the throat


In this article we analyze various properties of the charged traversable wormhole solution derived by Kim and Lee\cite{kim}. In particular, our main goal is to compute the shadow of the wormhole, which is relevant from an astrophysical point of view. Though shadow of wormholes were calculated for various wormhole geometries \cite{javed,reg,an}, Kim and Lee's solution has not been explored in the literature yet.

We first provide a short review of the main properties of the charged transversable wormhole. Then, by solving the geodesic equations for null particles, we determine the location of the photon ring (Section \ref{geo:photons}). We also study the trajectories of massive particles (Section \ref{geo:massive}) and in Section \ref{geo:gene} we extend these calculations to a more general charged wormhole geometry. Finally, we present in Section \ref{shadow} the images of the shadow of the wormhole when the light source is a star located beyond the mouth opposite to the observer's. We produce images for several values of the charge and viewing angles. The final section of the article is devoted to the conclusions. Throughout this work, we employ geometrized units $G = c = 1$.

\section{Charged Traversable Wormholes}

The solution for a charged traversable wormhole was found by Kim and Lee \cite{kim} and was generalized in general relativity and modified gravity \cite{kuf,sah,godsam}. The line element is given by:

\begin{equation}
	ds^{2} = -\Bigg(1+\frac{Q^{2}}{r^{2}}\Bigg)dt^{2} + \frac{1}{1-\frac{b(r)}{r}+\frac{Q^{2}}{r^{2}}}dr^{2} + r^{2}\big(d\theta^{2}+ sin^{2}(\theta) d\phi^{2}\big)\,.
	\label{eq31}
\end{equation}

		
		The corresponding components of the energy-momentum tensor\footnote{A didactic derivation was presented by Kimet \cite{kimet}} are \cite{kim}, .
		\begin{equation}
			\rho(r) = \frac{1}{8\pi}\Bigg\{\frac{b'(r)}{r^{2}} - \frac{Q^2}{r^{4}} \Bigg\}\,.
			\label{eq34}
		\end{equation}
		\begin{equation}
			\sigma(r) = -\frac{1}{8\pi}\Bigg\{\frac{b(r)}{r^{3}} - 2\Bigg(1-\frac{b(r)}{r}\Bigg)\frac{\Phi '(r)}{r} + \frac{Q^2}{r^{4}}\Bigg\}\,.
			\label{eq35}
		\end{equation}
		$$p(r) = \frac{1}{8\pi}\Bigg\{\Bigg(1-\frac{b(r)}{r}\Bigg)\Bigg[ \Phi''(r) + (\Phi'(r))^{2} - \frac{b'(r)r-b(r)}{2r(r-b(r))}\Phi'(r)-\frac{b'(r)r-b(r)}{2r^{2}(r-b(r))}+$$
		\begin{equation}
			+\frac{\Phi'(r)}{r} - \frac{Q^2}{r^{4}} \Bigg] \Bigg\}\,.
			\label{eq36}
		\end{equation}
		Where $\rho(r)$ is the energy density, $\sigma(r)$ is the tension and $p(r)$ is the pressure. In the present work, we will focus on the shape function:
		\begin{equation}
			b(r) = \frac{b_{0}^2}{r}\,.
			\label{eq37}
		\end{equation}
		When $Q$ = 0, the solution describes a neutral traversable wormhole, and when $b(r) = 0$ the spacetime geometry corresponds to the massless Reissner-Norström solution.
		
		The throat of the wormhole is located at 
		\begin{equation}
			1 - \frac{b^2_0}{r^2}+\frac{Q^{2}}{r^{2}} = 0,    
		\end{equation}
		that is
		\begin{equation}\label{throat}
			r = \pm \sqrt{b^2_0 - Q^2}.
		\end{equation}
		Notice that the coordinate $r$ is defined in the range $r \in (-\infty, - \sqrt{b^2_0 - Q^2}] \cup [+ \sqrt{b^2_0 - Q^2}, + \infty)$. 
		
		
		
		\subsection{\textbf{Geodesics} of Charged Traversable Wormholes}
		
		\subsubsection{\textbf{Trajectories of photons}}\label{geo:photons}
		
		We begin by computing the trajectories of null particles. From the line element  $(31)$, the Lagrangian of the charged wormhole is
		
		\begin{equation}
			2\mathcal{L} = -\Bigg(1+\frac{Q^{2}}{r^{2}}\Bigg)\dot{t}^{2} + \frac{1}{1-\frac{b_{0}^2}{r^2}+\frac{Q^{2}}{r^{2}}}\dot{r}^{2} + r^{2}\dot{\theta}^{2}+ r^{2}sin^{2}(\theta)\dot{\phi}^{2}\,,
			\label{eq39}
		\end{equation}
		where the dot is a derivative with respect to the affine parameter $\lambda$: $\dot{x^{\alpha}} =: \mathrm{d}x^{\alpha}/\mathrm{d}\lambda$. Since the metric coefficients are independent of $t$ and $\phi$, the Killing vectors of the spacetime are $k^{0} = (1,0,0,0)$ and $k^{3} = (0,0,0,1)$. Therefore, the constants of motion for $t$ direction and $\phi$ direction are: 
		\begin{equation}
			k^{0}u_{0} = -E\,,
			\label{eq40}
		\end{equation}
		
		\begin{equation}
			k^{3}u_{3} = \ell\,.
			\label{eq41}
		\end{equation}
		where $u_{\mu} = g_{\mu\nu}u^{\nu}$. The geodesic equations for $t$ coordinate and $\phi$ coordinate now take the simple form
		\begin{equation}
			\frac{\mathrm{d}t}{\mathrm{d}\lambda} = u^{0} = g^{0\nu}u_{\nu} = g^{00}u_{0} = \frac{E}{1+\frac{Q^2}{r^2}}\,,
			\label{eq42}
		\end{equation}
		
		\begin{equation}
			\frac{\mathrm{d}\phi}{\mathrm{d}\lambda} = u^{3} = g^{3\nu}u_{\nu} = g^{33}u_{3} = \frac{\ell}{r^2sin^2(\theta)}\,,
			\label{eq43}
		\end{equation}
		where $E$ and $\ell$ are identified with the energy and angular momentum, respectively. The geodesic equations for the $r$ and $\theta$ components will be calculated using the Hamilton-Jacobi approach \cite{ferrari} as follows
		\begin{equation}
			\frac{\partial S}{\partial \lambda} + \frac{1}{2}g^{\mu\nu}\frac{\partial S}{\partial x^{\mu}}\frac{\partial S}{\partial x^\nu} = 0\,,
			\label{eq44}
		\end{equation}
		or equivalently \cite{ferrari}
		\begin{equation}
			-\kappa + g^{\mu\nu}\frac{\partial S}{\partial x^{\mu}}\frac{\partial S}{\partial x^\nu} = 0\,.
			\label{eq45}
		\end{equation}
		We assume the following ansatz for the $S$ function:
		\begin{equation}
			S = -\frac{1}{2}\kappa \lambda - Et + \ell \phi + S^{(r)}(r) + S^{(\theta)}(\theta)\,,
			\label{eq46}
		\end{equation}
		where in the case of null paths, the constant $\kappa$ in both \ref{eq45} and \ref{eq46} is set to zero. Replacing \ref{eq46} into \ref{eq45} we obtain
		
		$$  g^{\mu\nu}\frac{\partial S}{\partial x^{\mu}}\frac{\partial S}{\partial x^\nu} = g^{00}\Bigg(\frac{\partial S}{\partial x^{0}}\Bigg)^{2} + g^{11}\Bigg(\frac{\partial S}{\partial x^{1}}\Bigg)^{2} + g^{22}\Bigg(\frac{\partial S}{\partial x^{2}}\Bigg)^{2}+ g^{33}\Bigg(\frac{\partial S}{\partial x^{3}}\Bigg)^{2} =$$
		
		\begin{equation}
			\frac{-E^2}{\Big(1+\frac{Q^2}{r^2}\Big)} + \Bigg(1-\frac{b_{0}^2}{r^{2}}+\frac{Q^2}{r^2}\Bigg)\Bigg(\frac{\partial S^{(r)}(r)}{\partial r}\Bigg)^{2}+\frac{1}{r^2}\Bigg(\frac{\partial S^{(\theta)}(\theta)}{\partial \theta}\Bigg)^{2}+\frac{\ell^{2}}{r^{2}sin^{2}(\theta)} = 0\,.
			\label{eq47}
		\end{equation}
		Multiplying both sides by a $r^{2}$, the latter equation becomes
		\begin{equation}
			\frac{-r^2 E^2}{\Big(1+\frac{Q^2}{r^2}\Big)} + r^2\Bigg(1-\frac{b_{0}^2}{r^{2}}+\frac{Q^2}{r^2}\Bigg)\Bigg(\frac{\partial S^{(r)}(r)}{\partial r}\Bigg)^{2}+\Bigg(\frac{\partial S^{(\theta)}(\theta)}{\partial \theta}\Bigg)^{2}+\frac{\ell^{2}}{sin^{2}(\theta)} = 0\,.
			\label{eq48}
		\end{equation}
		The equation can be separated as:
		\begin{equation}
			\frac{r^2 E^2}{\Big(1+\frac{Q^2}{r^2}\Big)} - r^2\Bigg(1-\frac{b_{0}^2}{r^{2}}+\frac{Q^2}{r^2}\Bigg)\Bigg(\frac{\partial S^{(r)}(r)}{\partial r}\Bigg)^{2} = \Bigg(\frac{\partial S^{(\theta)}(\theta)}{\partial \theta}\Bigg)^{2}+\frac{\ell^{2}}{sin^{2}(\theta)} = \mathcal{K}\,,
			\label{eq49}
		\end{equation}
		where $\mathcal{K}$ is the Carter constant \cite{chandra,carter}. Now we have two equations
		\begin{equation}
			\begin{cases}
				\Bigg(1-\frac{b_{0}^2}{r^{2}}+\frac{Q^2}{r^2}\Bigg)\Bigg(\frac{\partial S^{(r)}(r)}{\partial r}\Bigg)^{2} = \frac{E^2}{\Big(1+\frac{Q^2}{r^2}\Big)} - \frac{\mathcal{K}}{r^2}  \\ \Bigg(\frac{\partial S^{(\theta)}(\theta)}{\partial \theta}\Bigg)^{2} = \mathcal{K} - \frac{\ell^{2}}{sin^{2}(\theta)}
			\end{cases}\,.
			\label{eq50}
		\end{equation}
		Using now the relations\cite{ferrari}:
		\begin{equation}
			p_{r} = \frac{\partial \mathcal{L}}{\partial \dot{r}} = \frac{\partial S^{(r)}(r)}{\partial r} = \frac{\dot{r}}{\Bigg(1-\frac{b_{0}^2}{r^{2}}+\frac{Q^2}{r^2}\Bigg)}\,,
			\label{eq51}
		\end{equation}
		
		\begin{equation}
			p_{\theta} = \frac{\partial \mathcal{L}}{\partial \dot{\theta}} = \frac{\partial S^{(\theta)}(\theta)}{\partial \theta} =r^2\dot{\theta}\,,
			\label{eq52}
		\end{equation}
		The set of  equations in \ref{eq50} becomes
		\begin{equation}
			\begin{cases}
				\frac{\dot{r}}{\sqrt{1-\frac{b_{0}^2}{r^2}+\frac{Q^2}{r^2}}}  = \sqrt{\frac{E^2}{1+\frac{Q^2}{r^2}}-\frac{\mathcal{K}}{r^2}} \\   r^{2}\dot{\theta} = \sqrt{\mathcal{K}-\frac{\ell ^{2}}{sin^{2}\theta}}
			\end{cases}\,,
			\label{eq53}
		\end{equation}
		
		The complete set of null geodesic equations is
		
		\begin{equation}
			\Bigg(1+\frac{Q^2}{r^2}\Bigg)\frac{\mathrm{d}t}{\mathrm{d}\lambda} = E\,,
			\label{eq54}
		\end{equation}
		
		\begin{equation}
			\frac{1}{\sqrt{1-\frac{b_{0}^2}{r^2}+\frac{Q^2}{r^2}}} \frac{\mathrm{d}r}{\mathrm{d}\lambda} = \sqrt{\frac{E^2}{1+\frac{Q^2}{r^2}}-\frac{\mathcal{K}}{r^2}}\,,
			\label{eq55}
		\end{equation}
		
		\begin{equation}
			r^{2}\frac{\mathrm{d}\theta}{\mathrm{d}\lambda} = \sqrt{\mathcal{K}-\frac{\ell ^{2}}{sin^{2}\theta}}\,,
			\label{eq56}
		\end{equation}
		
		\begin{equation}
			\frac{\mathrm{d}\phi}{\mathrm{d}\lambda} = \frac{\ell}{r^{2}sin^{2}\theta}\,.
			\label{eq57}
		\end{equation}
		
		The $r$ and $\theta$ geodesic equations can be rewritten as
		\begin{equation}
			\frac{1}{\sqrt{1-\frac{b_{0}^2}{r^2}+\frac{Q^2}{r^2}}} \frac{\mathrm{d}r}{\mathrm{d}\lambda} = \pm\sqrt{A(r)}\,,
			\label{eq58}
		\end{equation}
		
		\begin{equation}
			r^{2}\frac{\mathrm{d}\theta}{\mathrm{d}\lambda} = \pm\sqrt{T(\theta)}\,.
			\label{eq59}
		\end{equation}
		where
		\begin{equation}
			A(r) = E^{2
			}\Bigg(\frac{1}{1+\frac{Q^2}{r^2}}-\frac{\mathcal{\eta}}{r^2}\Bigg)\,,
			\label{eq61}
		\end{equation}
		
		\begin{equation}
			r^{2}\frac{\mathrm{d}\theta}{\mathrm{d}\lambda} = \pm E\sqrt{\eta-\frac{\xi ^{2}}{sin^{2}\theta}}\,,
			\label{eq62}
		\end{equation}
		and
		\begin{equation}
			\begin{cases}
				\xi = \frac{\ell}{E}, \\ \eta = \frac{\mathcal{K}}{E^2}.
			\end{cases}\,
			\label{eq60}
		\end{equation}
		
		An effective potential can be defined as follows

		$$\frac{\mathrm{d}r}{\mathrm{d}\lambda} = \sqrt{A(r)}\sqrt{1-\frac{b_{0}^2}{r^2}+\frac{Q^2}{r^2}} \iff  \Big(\frac{\mathrm{d}r}{\mathrm{d}\lambda}\Big)^{2} = \Bigg(1-\frac{b_{0}^2}{r^2}+\frac{Q^2}{r^2}\Bigg)A(r) \iff$$
		
		$$ \iff \Big(\frac{\mathrm{d}r}{\mathrm{d}\lambda}\Big)^{2} - \Bigg(1-\frac{b_{0}^2}{r^2}+\frac{Q^2}{r^2}\Bigg)A(r) = 0 \implies$$
		
		\begin{equation}
			\Big(\frac{\mathrm{d}r}{\mathrm{d}\lambda}\Big)^{2} + V_{\mathrm{eff}} = 0\,,
			\label{eq63} 
		\end{equation}
		
		where $V_{\mathrm{eff}}$ is:
		
		\begin{equation}
			V_{\mathrm{eff}} =:  -\Bigg(1-\frac{b_{0}^2}{r^2}+\frac{Q^2}{r^2}\Bigg)A(r)\,.
			\label{eq64}
		\end{equation}
		Equation \ref{eq64} implies: 
		\begin{equation}
			A(r) = 0\,,
			\label{eq65}
		\end{equation}
		and
		\begin{equation}
			A'(r) = 0\,.
			\label{eq66}
		\end{equation}
		Condition \ref{eq65} allows to fix the $\eta$ parameter
		\begin{equation}
			\eta = \frac{r^2}{1+\frac{Q^2}{r^2}}\,,
			\label{eq67}
		\end{equation}
		
		We derive the radius of the unstable circular photon orbit, $r_{ph}$, from the following conditions on the effective potential:
		
		\begin{equation}
			V_{\mathrm{eff}}\Big \|_{r=r_{ph}}=0, \hspace{0.5cm} \frac{\mathrm{d}}{\mathrm{d}r}V_{\mathrm{eff}}\Big \|_{r=r_{ph}}=0\hspace{0.5cm} \frac{\mathrm{d}^{2}}{\mathrm{d}r^{2}}V_{\mathrm{eff}}\Big \|_{r=r_{ph}}<0\,.
			\label{eq68}
		\end{equation}
		
		The first derivative reads:
		
		$$E^2 \left(\frac{b_{0}^2}{r_{ph}^2}-\frac{Q^2}{r_{ph}^2}-1\right) \left(\frac{2 Q^2}{r_{ph}^3 \left(\frac{Q^2}{r_{ph}^2}+1\right)^2}+\frac{2 \eta }{r_{ph}^3}\right)+$$
		
		\begin{equation}        
			+E ^2 \left(\frac{2
				Q^2}{r_{ph}^3}-\frac{2 b_{0}^2}{r_{ph}^3}\right) \left(\frac{1}{\frac{Q^2}{r_{ph}^2}+1}-\frac{\eta }{r_{ph}^2}\right)=0\,.
			\label{eq69}
		\end{equation}
		
		Substitution of \ref{eq67} into \ref{eq69} yields:
		
		\begin{equation}        
			-\frac{2 E^2 \left(2 Q^2+r_{ph}^2\right) \left(-b_{0}^2+Q^2+r_{ph}^2\right)}{r_{ph} \left(Q^2+r_{ph}^2\right)^2}=0\,,
			\label{eq70}
		\end{equation}
		whose solution is 
		\begin{equation}
			r_{ph} = \pm\sqrt{b_{0}^2-Q^{2}}\,.
			\label{eq71}
		\end{equation}
		
		The radius of the unstable photon orbit therefore is a function of the charge of the wormholes and the constant $b_0$. There is one unstable circular orbit at each side of the throat. Notice that
		\begin{equation}
			b_{0}>Q\,,
			\label{eq72}
		\end{equation}
		which agrees with the condition found in\cite{kim}.
		
		Notice that the photon orbit coincides with the wormhole's throat (see Eq. \ref{throat}).
		
		\subsection{The photon ring of a charged traversable wormhole}
		
\begin{figure}[H]
			\centerline{\includegraphics[width=12cm]{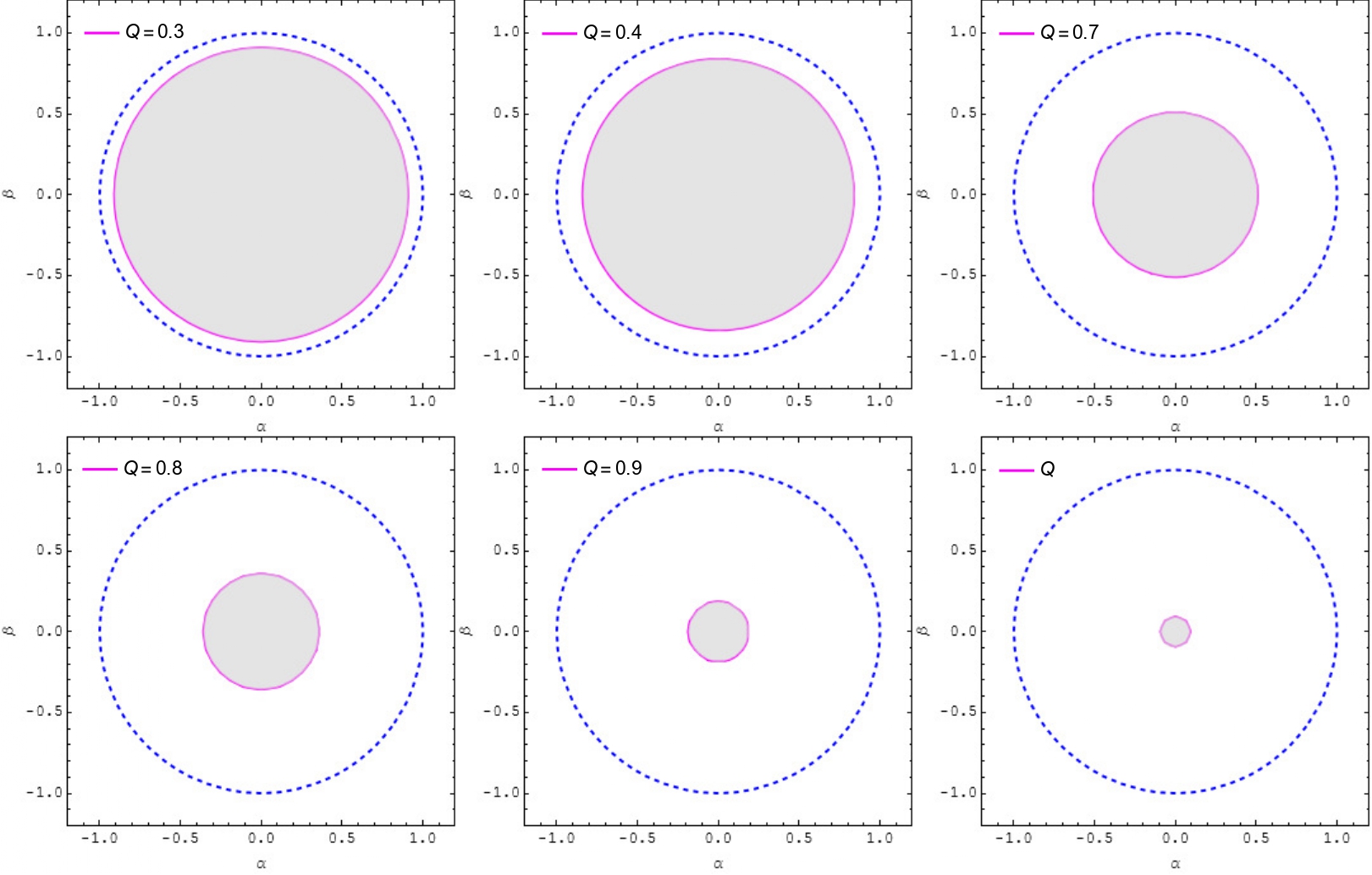}}
			\caption{Photon ring of the CTW for different values of the charge $Q$ and radius $b_{0} = 1$. The dashed circle corresponds to $Q = 0$.} \label{f1}
		\end{figure}

		Here we show how the photon ring is projected in the sky. To that end, we introduce celestial coordinates that are defined as \cite{chandra}:
		\begin{equation}
			\alpha = \lim _{r_{0} \to \infty} \Bigg(-r_{0}^2 sin\theta_{0}\frac{\mathrm{d}\phi}{\mathrm{d}r}\Bigg)\,,
			\label{eq73}
		\end{equation}
		and
		\begin{equation}
			\beta = \lim _{r_{0} \to \infty} \Bigg(r^{2}_{0}\frac{\mathrm{d}\theta}{\mathrm{d}r} \Bigg)\,,
			\label{eq74}
		\end{equation}
		where $r_{0}$ is the position coordinate of the remote observer and $\theta_{0}$ the angle of inclination between the axis of symmetry of the wormhole and the
		direction to the far distant observer. Using the geodesic equations, we get 
		
		
		\begin{equation}
			\frac{\mathrm{d}\phi}{\mathrm{d}r} = \frac{\xi  \csc ^2(\theta )}{r^2 E \sqrt{1-\frac{b_{0}^2}{r^2}+\frac{Q^2}{r^2}} \sqrt{\frac{1}{\frac{Q^2}{r^2}+1}-\frac{\eta
					}{r^2}}}\,,
			\label{eq78}
		\end{equation}
		\begin{equation}
			\frac{\mathrm{d}\theta}{\mathrm{d}r} = \frac{\sqrt{\eta -\xi ^2 \csc ^2(\theta )}}{r^2 E \sqrt{1-\frac{b_{0}^2}{r^2}+\frac{Q^2}{r^2}} \sqrt{\frac{1}{\frac{Q^2}{r^2}+1}-\frac{\eta
					}{r^2}}}\,.
			\label{eq79}
		\end{equation}
		Then, expressions \ref{eq73} and \ref{eq74} become
		\begin{equation}
			\alpha = \lim _{r_{0} \to \infty} \Bigg(-\frac{\xi  \csc (\theta_{0} )}{\sqrt{1-\frac{b_{0}^2}{r_{0}^2}+\frac{Q^2}{r_{0}^2}}
				\sqrt{\frac{1}{\frac{Q^2}{r_{0}^2}+1}-\frac{\eta }{r_{0}^2}}}\Bigg)\,,
			\label{eq80}
		\end{equation}   
		\begin{equation}
			\beta = \lim _{r_{0} \to \infty} \Bigg( \frac{\sqrt{\eta -\xi ^2 \csc ^2(\theta_{0}
					)}}{\sqrt{1-\frac{b_{0}^2}{r_{0}^2}+\frac{Q^2}{r_{0}^2}}
				\sqrt{\frac{1}{\frac{Q^2}{r_{0}^2}+1}-\frac{\eta }{r_{0}^2}}}\Bigg)\,.
			\label{eq81}
		\end{equation} 
		After some algebraic simplifications, we obtain
		\begin{equation}
			\alpha = - \xi \mathrm{csc}(\theta_{0})\,,
			\label{eq82}
		\end{equation}
		\begin{equation}
			\beta = \sqrt{\eta -\xi ^2 \mathrm{csc} ^2(\theta_{0} )}\,.
			\label{eq83}
		\end{equation}
		Finally, the photon ring $ \mathcal{S}$  of the CTW, is the geometrical locus given by
		\begin{equation}
			\mathcal{S}(\alpha,\beta) =: \alpha^{2} + \beta^{2}\,,
			\label{eq84}.
		\end{equation}
		and
		\begin{equation}
			\mathcal{S}(\alpha,\beta) = \eta = \frac{r_{ph}^{2}}{1+\frac{Q^{2}}{r_{ph}^{2}}} = \frac{\big(b_{0}^{2}-Q^{2}\big)^{2}}{b_{0}^{2}}\,.
			\label{eq85}
		\end{equation}
		We show in Figure \ref{f1} the projection of the photon ring in the sky. We see that as the value of the charge increases, the size of the photon ring decreases, i.e. the wormhole becomes more compact.
		
		
		\subsection{Trajectories of massive neutral particles}\label{geo:massive}
		
		Here, we show that the trajectories of massive neutral particles have always a radial positive acceleration; thus, in absence of additional forces, particles can only but escape from the wormhole.
		
		The Lagrange equations for  $t$ component and $\phi$ component are
		\begin{eqnarray}
			-\left(1+\frac{Q^{2}}{r^{2}}\right) \dot{t} & = & k, \label{k} \\
			r^2 \sin^2 \theta \dot{\phi} & = & h,\label{h}
		\end{eqnarray}
		where $k$ and $h$ are constants. Since the metric is spherically symmetric, we can study without loss of generality, the motion of geodesics in the equatorial plane ($\theta = \pi/2 $). The equation of motion for the radial component is
		\begin{equation}\label{accel}
			\ddot{r} = \frac{\dot{r}^2}{r^{3}}\frac{b_{0}^2 -Q^2}{{1-\frac{b_{0}^2}{r^2}+\frac{Q^2}{r^2}}} + \left(1-\frac{b_{0}^2}{r^2}+\frac{Q^2}{r^2}\right)\frac{1}{r^3} \left[\frac{Q^2 k^2}{\left(1+\frac{Q^2}{r^2}\right)^2}+ h^2\right]. 
		\end{equation}
		In the latter equation, we employ expressions \ref{k} and \ref{h}. Since $b_0 > Q$ (this condition is required in order to maintain the wormhole under the addition of electric charge\cite{kim+01}) and
		\begin{equation}
			1 - \frac{b_0^2}{r^{2}} - \frac{Q^2}{r^2} > 0,    
		\end{equation}
		for $r > \sqrt{b_0^2 - Q^2}$ (spacetime region outside the throat), the right-hand side of Eq. \eqref{accel} is always positive and any massive particle is accelerated in the opposite direction of the wormhole. We see that this charged wormhole geometry exerts gravitational repulsion on material neutral particles.
		
		We have found that this particular feature is also present in a more general class of charged wormholes geometries, as we show next.
		
		\subsubsection{Generalized charged wormhole geometry}\label{geo:gene}
		
		The spacetime metric given by Eq. \ref{eq31} can be generalized as follows\cite{kuf}
		\begin{equation}
			ds^{2} = - \left(1+ R(r) + \frac{Q^2}{r^{2}}\right) dt^{2} + \left(1 - \frac{b(r)}{r^{2}} + \frac{Q^2}{r^{2}}\right)^{-1} dr^{2} + r^{2} \left(d\theta^{2} + \sin^2{\theta} d\phi^{2}\right),   
		\end{equation}
		where the function $R(r) \ge 0$ to avoid the presence of an event horizon. Additionally, $R(r)$ should have a continuous derivative. Asymptotic flatness requires that
		\begin{eqnarray}
			\lim_{r \to +\infty} R(r) & = & 0,\\
			\lim_{r \to +\infty} R'(r) & = & 0.
		\end{eqnarray}
		From the Euler-Lagrange equations, we compute the radial component of the equation. It reads,
		\begin{equation}
			\ddot{r} = \frac{\dot{r}^2}{2}\frac{\left(\frac{b(r)}{r^2} -\frac{b'(r)}{r} - \frac{2Q^2}{r^3}\right)}{1- \frac{b(r)}{r} + \frac{Q^2}{r^2}} + \left(1- \frac{b(r)}{r} + \frac{Q^2}{r^2}\right)\left[\frac{1}{2} \frac{\left(-R'(r)+2 \frac{ Q^2}{r^3}\right)}{\left(1+R(r)+\frac{Q^2}{r^2}\right)^2} + \frac{h^2}{r^3}\right].
		\end{equation}
		If we assume that\footnote{Notice that this particular prescription satisfies the conditions for asymptotic flatness.}
		\begin{equation}
			R(r) = b(r) = \frac{b^2_0}{r},    
		\end{equation}
		then
		\begin{eqnarray}
			\frac{b(r)}{r^2} -\frac{b'(r)}{r} - \frac{2Q^2}{r^3} & = & \frac{2}{r^3} b^2_0 -Q^2 > 0,\\
			1- \frac{b(r)}{r} + \frac{Q^2}{r^2} & = & 1 - \frac{b_0^2}{r^{2}} - \frac{Q^2}{r^2} > 0,\\
			-R'(r)+2 \frac{ Q^2}{r^3} & = & \frac{b^2_0}{r^2} + 2 \frac{ Q^2}{r^3} > 0.
		\end{eqnarray}
		Consequently, the radial acceleration $\ddot{r} > 0$: the charged wormhole exerts gravitational repulsion on free neutral particles.
		
		When we think of accretion processes, gravity plays the role of `attracting'' matter in such a way that different structures are formed. In our case, the wormhole geometry acts as a repellent, impeding  the accumulation of matter and the formation of well-known structures such as accretion disks.
		
		Since the goal of this work is to compute and analyze the properties of the shadow of this particular wormhole geometry, we need a source of photons whose trajectories integrated in a region of spacetime will provide the image or ``shadow'' seen by an observer at a certain distance. From the present discussion, it is apparent that the light source could not come form an accretion disk or some other structure around the wormhole. Instead, we will consider as the light source a star that is in the other side of the throat, so that observer and star are in different spacetime regions. The wormhole is orbiting the star. An observer in the other side of the mouth, in principle, will be able to see the star. In the next section, we compute the corresponding shadow for this particular configuration.

\section{Wormhole shadow: the image of a star in the other side of the throat}\label{shadow}

We use the code \texttt{SKYLIGHT} \cite{pelle2022skylight} to produce the images of the wormhole shadow. \texttt{SKYLIGHT} is a general-relativistic ray tracing and radiative transfer code for arbitrary spacetimes and coordinate systems. In particular, we employ the observer-to-emitter scheme of the code, in which a virtual image plane is set at the location of the observer, and the spacetime geodesics are traced backwards in time from the center of each image pixel towards the light source. (See Fig. 1 of the \texttt{SKYLIGHT}'s reference for clarity.) In the code, the geodesics are numerically integrated starting from the observer until a halting condition is met. For this setting, we have two halting conditions: either (a) the geodesic intersects the emitting surface, or (b) the geodesic goes far enough away to ensure that it does not intersect the emitting surface. It is worth noticing that condition (b) might be met in either side of the throat. 

Since the star is located in the other side of the throat, we use the $\ell$-coordinate\footnote{The coordinate is defined as \begin{equation}
		\ell (r) = \pm \int^{r}_{r_{0}} \Bigg(1- \frac{b(r)}{r} \Bigg)^{-\frac{1}{2}} dr^{2}\,.
		\label{eq11} 
\end{equation}} for the numerical integration instead of $r$. This is because the former coordinate system is globally regular, while the latter has an apparent singularity at the wormhole's throat. In our algorithm, the center of the image plane (where the observer is located) is at $\ell = 500 b_0$ and $\phi=0$. The $\theta$ coordinate of the center of the image plane is the viewing angle, and we let it take different values. In what follows, we denote the viewing angle $\xi$. In this work, we model the surface of the star as the locus of the following equations:
\begin{align}
	(x-x_c)^2 + y^2 + z^2 &= R^2\,, \\
	\ell < 0 \,,
	\label{eq:star_surface}
\end{align}
where
\begin{align}
	x &= r(\ell) \sin \theta \cos \phi\,, \\
	y &= r(\ell) \sin \theta \sin \phi\,, \\
	z &= r(\ell) \cos \theta \,,
\end{align}
and
\begin{align}
	x_c &= r(\ell_c)\,, \\
	\ell_c &= -10 b_0\,, \\
	R &= 5 b_0\,.
\end{align}
Here, $R$ is a parameter of the model related to the surface of the star.

\begin{figure}
	\centering
	\includegraphics[width = 10cm]{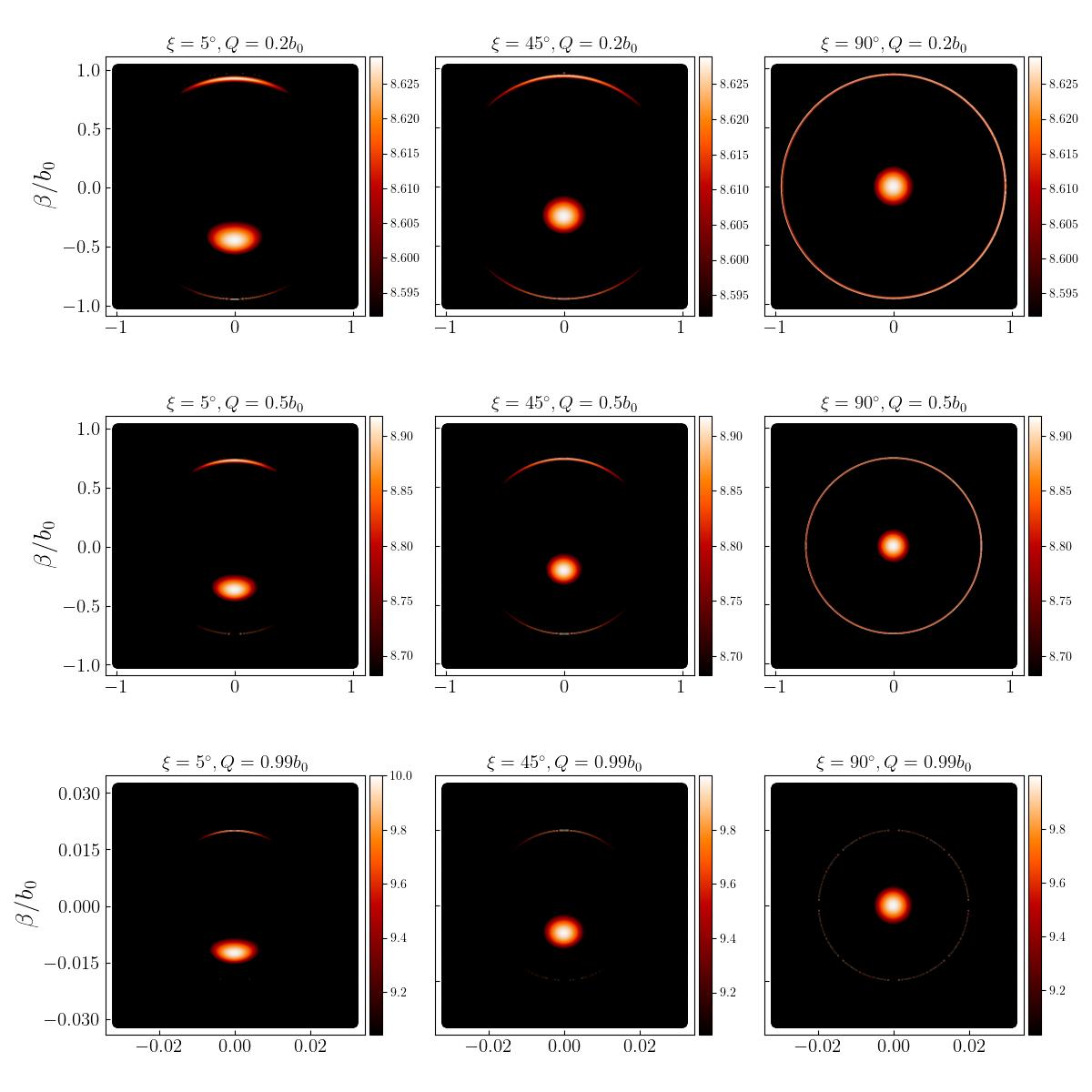}
	\caption{Bolometric images of a star of uniform temperature in the opposite side of a wormhole's throat, for charges $Q/b_0= 0.2$, $0.5$, $0.99$, and viewing angles $\xi = 5^\circ$, $45^\circ$, $90^\circ$. $\alpha$ and $\beta$ are rectangular coordinates over the image plane. The units of flux are arbitrary (as is the temperature of the star).}
	\label{fig:fluxes}
\end{figure}

We assume the star has a uniform superficial temperature. Only those rays which intersect the surface contribute to the image.  The observed bolometric flux at each of these pixels is 
\begin{equation}
	F = (1+z)^4 \sigma_B T^4\,,
\end{equation}
where $z$ is the redshift of the ray along its path between the observer and the source, $\sigma_B$ is the Stefan-Boltzmann constant, and $T$ is the temperature of the star. In Fig.\ref{fig:fluxes}, we show a set of bolometric images for three different values of the charge $Q/b_0= 0.2$, $0.5$, $0.99$, and  viewing angles, $\xi = 5^\circ$, $45^\circ$, $90^\circ$. In the plots, $\alpha$ and $\beta$ are rectangular coordinates over the image plane.

We first notice that light rays emitted from the star reach the observer independently of the viewing angle, even in the case where the observer is located almost on the $z$-axis ($\xi = 5^\circ$). When the observer and the star are perfectly aligned ($\xi = 90^\circ$), we can clearly identify two disconnected images: a primary image, composed by light rays that cross the throat and reach the observer directly; and a secondary image, corresponding to light rays which pass almost tangent to the throat surface. Since the throat is a null surface, light rays almost tangent to it are subject to extreme gravitational deflection. The secondary image is the Einstein ring. 

\begin{figure}
	\centering
	\includegraphics[width = 10cm]{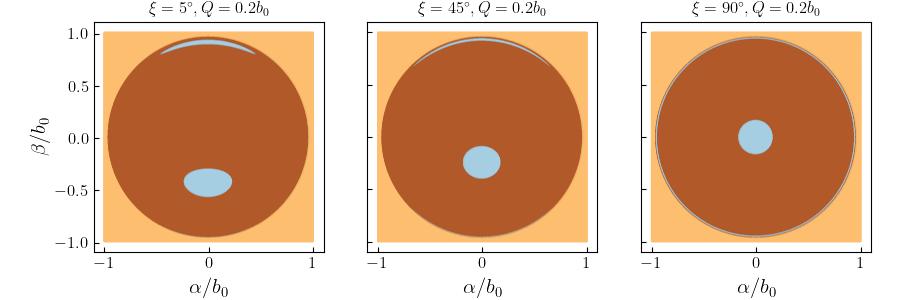}
	\caption{State of the light rays after numerical integration for charge $Q = 0.2b_0$ and viewing angles $\xi = 5^\circ$, $45^\circ$, $90^\circ$. The color is set as follows: a) in cyan, rays that intersect the surface of the star, b) in light brown, rays which go to infinity on the same asymptotic side as the observer, and (c) in dark brown, rays that cross the wormhole's throat and go to infinity in the opposite side of the observer, i.e. on the same side as the star. $\alpha$ and $\beta$ are rectangular coordinates over the image plane.}
	\label{fig:endstaes}
\end{figure}

To illustrate the apparent location of the throat as seen by the observer, in Fig.\ref{fig:endstaes} we classify the behavior of light rays according to their state after a halting condition for the numerical integration is met: a) the region in cyan corresponds to those rays that intersect the surface of the star, b) in light brown, those rays which go to infinity on the same asymptotic side as the observer, and (c) in dark brown, those rays which go to infinity on the other asymptotic side (i.e. after traversing the wormhole). These plots allow us to identify the profile of the wormhole's throat (the Einstein ring for $\xi = 90^\circ$) as the boundary between the light and dark brown regions. The results here obtained are in agreement with the detailed analysis by Rajbul et al. \cite{raj+19}; they computed the strong gravitational lensing by generic spherically symmetric static wormholes taking into account the case where the photon ring coincides with the throat and the light source is on the opposite side of the observer.

We have also found that the apparent size of the throat decreases when the spacetime charge increases, which is consistent with the results described in the previous sections; furthermore, the structure of the images is preserved among the different values of the charge, even though the wormhole throat shrinks. This happens despite the fact that the ``physical" size of the star is the same size in all cases. When the observer is on the $z$-axis, our ``Euclidean" intuition could deceive us and let us think that we should not receive any light rays from the star. However, the wormhole geometry deflects photons in such a way that some of them are able to reach the observer.  


The images here presented bear some similarities (and also some differences) with those of some gravastars\footnote{Gravastars are supercompact objects which result from the gravitational collapse of stars. Their external geometry is that of a black hole but they lack event horizons. These models are also singularity free.} models. Sakai and coworkers \cite{sak+14} (see also the work by Kubo and Sakai\cite{kub+16}) computed the shadow of a gravastar model developed by Visser and Wilshire \cite{vis+04} assuming that the surface of the gravastar is electromagnetically transparent. They consider a star rotating around the gravastar.

When the star is behind the gravastar, as seen by a given observer, the image produced is composed by a small disk and arcs around it, as in this work. Notice that the disk would be absent if the central object were a black hole. The external geometry of this particular gravastar is the Schwarzschild metric, so  the image has an additional set of outer light arcs that are produced by those photons that do not penetrate inside the gravastar. Naturally, this feature is absent in the charged wormhole shadow. Though wormholes and gravastars represent completely different geometries, their shadows look quite alike since both lack event horizons and singularities.



\section{Conclusion}\label{conclusion}

In this work we have made a comprehensive analysis of the properties of a charged wormhole spacetime \cite{kim}. In particular, we have focused on a traversable wormhole solution with a shape function given by $b(r)=b_{0}^2/r$. We have computed the spacetime geodesic equations for null particles and determined that there is just one equatorial circular photon orbit, which is unstable, and is located at the throat. We have found that the radius of the photon ring decreases when the value of the charge increases.

We also computed the equation of motion for time-like trajectories and found out that the charged wormhole exerts gravitational repulsion on free neutral particles, i.e. the radial acceleration of these particles is always positive. Even for more general charged wormhole geometries, this feature is also present. Consequently, it is not possible, in principle, the formation of structures by accretion processes around the wormhole.

Given these constraints, a possible scenario where an observer could detect the shadow of the wormhole is assuming that the wormhole is orbiting a star which is located in the opposite mouth of the observer. Some light rays from the star will go through the throat and, depending on their impact parameter, will reach the observer.

We obtained the images of the wormhole shadow by using \texttt{SKYLIGHT}, a general-relativistic ray tracing and radiative transfer code for arbitrary spacetimes and coordinate systems that was developed by Pelle and collaborators \cite{pelle2022skylight}. The most relevant features we encountered are: light rays from the star reach the observer for all viewing angles; in all cases, a secondary image is visible due to light rays that pass almost tangent to the throat, which is a null surface; when the observer and the star are perfectly aligned, an Einstein ring forms at the apparent location of the throat. In accordance with our previous analysis, when the charge increases the size of the apparent size of the throat decreases and the size of the image scales proportionally. The structure of the image depends only on the viewing angle but not on the charge.

This work is a first step towards more realistic models for the computation of the shadow of the charged wormhole. For instance, the star could have a non-uniform intensity profile that includes ``limb darking'' effects. We shall explore these issues in future works. 


\section*{Acknowledgments}


MRN would like to thank his advisor, Prof. Dr. Luiz Claudio Lima Botti, for the support on scientific research and friendship; the Instituto Nacional de Pesquisas Espaciais ($\mathrm{INPE}$), Universidade Federal de São Carlos ($\mathrm{UFSCar}$) and Centro de Radioastronomia e Astrofísica Mackenzie ($\mathrm{CRAAM}$). Furthermore, MRN would like to thank his family and friends. JP acknowledges Prof. Dr. Oscar Reula for very useful discussions. The authors are grateful to Prof. Dr. Gustavo E. Romero for his many insightful comments that helped to improve this article. 
MRN kindly acknowledges the received support from $\mathrm{CNPq}$ - $\mathrm{PIBIC/INPE}$ (process: $\mathrm{133024/2021-0}$). DP acknowledges the support from CONICET under grant PIP 0554. JP was supported by a CONICET fellowship. This work used computational resources from CCAD – Universidad Nacional de Córdoba (https://ccad.unc.edu.ar/), which are part of SNCAD – MinCyT, República Argentina.


\begin{thebibliography}{0}
\bibitem{visserb} M. Visser, {\it Lorentzian Wormholes}, 1st edn. (Springer-Verlag, New York, 1995).

\bibitem{flamm} L. Flamm, {\it Phy. Z.} {\bf 17}, 448 (1916).

\bibitem{weyl1} H. Weyl, {\it Philosophie der Mathematik und Naturwissenschaft}, (Leibniz Verlag, Munich, 1928) p.~65.

\bibitem{weyl2} H. Weyl, {\it Philosophy of Mathematics and Natural Science}, (Princeton University Press, Princeton, 1949) p.~91.

\bibitem{erb} A. Einstein and N. Rosen, {\it Phys. Rev.} {\bf 48} (1935).

\bibitem{wheeler55} J. A. Wheeler, {\it Phys. Rev.} {\bf 97}, 511 (1955).

\bibitem{wheeler+57} C. W. Misner and J. A. Wheeler, {\it Ann. Phys.} {\bf 2}, 525 (1957).

\bibitem{wheeler} J. A. Wheeler, {\it Geometrodynamics}, 1st edn. (Academic Press, New York and London, 1962).

\bibitem{fuller} R. W. Fuller and J. A. Wheeler, {\it Phys. Rev.} {\bf 128}, 2 (1962).

\bibitem{ellis1} H. G. Ellis, {\it Journal of Mathematical Physics} {\bf 14}, 1 (1973).

\bibitem{ellis2} H. G. Ellis, {\it General Relativity and Gravitation} {\bf 10}, 2 (1979).

\bibitem{bronn} K. A. Bronnikov, {\it Acta Phys. Polon. B.} {\bf 4},(1973).

\bibitem{mt} M.S. Morris and K. S. Throne, {\it Am. J. Phys.} {\bf 56}, 5 (1988).

\bibitem{cle1} G.Clément, {\it Am. J. Phys.} {\bf 57}, 967 (1989).

\bibitem{roman} T. A. Roman, {\it Phys. Rev. D.} {\bf 47}, 1370 (1993).

\bibitem{kim2} S. Kim, {\it Jour. Kor. Phys. Soc.} {\bf 63} (2013).

\bibitem{teo} E. Teo, {\it Phys. Rev. D.} {\bf 58}, 024014 (1998).

\bibitem{alias} A. Alias, I. Abdul and H.A. Kassim, Slowly Rotating Wormhole Solution, in {\it 1st International Conference on Mathematics, Statistics and Applications},(Lake Toba, Indonesia, 2005), v.{\bf 1}.

\bibitem{kim} S. Kim, H. Lee, {\it Phys. Rev. D.} {\bf 63}, 064014 (2001).

\bibitem{but} L.M. Butcher, {\it Phys. Rev. D.} {\bf 91}, 124031 (2015).

\bibitem{quinet} J.P.S. Lemos, F.S.N. Lobo, S.Q. Oliveira, {\it Phys. Rev. D.} {\bf 68}, 064004 (2003).

\bibitem{sant} A. C. L. Santos, C. R. Muniz, L.T. Oliveira, {\it Int. J. Mod. Phys. D.} {\bf 30}, 5 (2021).


\bibitem{dixit} A. Dixit , C. Chawla and A. Pradhan, {\it Int. J. Geom. Methods Mod. Phys.} {\bf 18}, 4 (2021).

\bibitem{mish} A. K. Mishra, U. K. Sharma, V. C. Dubey and A. Pradhan {\it Astro. Space. Sci.} {\bf 365}, 34 (2020)

\bibitem{bar} A. Baruah and A. Deshamukhya {\it IOP Conf. Journal of Physics} {\bf 1330}, 012001 (2019)


\bibitem{sharma} A. K. Mishra and U. K. Sharma, {\it New Astronomy} {\bf 88}, 101628 (2021).


\bibitem{vip} A. K. Mishra, V. C. Dubey and U. K. Sharma,, {\it Int. J. Geom. Methods Mod. Phys.} {\bf 17}, 2050155 (2020)

\bibitem{sam} G. C. Samanta , N. Godani and K. Bamba, {\it Int. J. Mod. Phys. D.} {\bf 29}, 9 (2021).

\bibitem{cat} M. Cataldo, L. Liempi and P. Rodrígues, {\it Eur. Phys. J. C.} {\bf 77}, 748 (2017).

\bibitem{god} N. Godani and G. C. Samanta, {\it Eur. Phys. J. C.} {\bf 80}, 30 (2020)

\bibitem{gau} N. Godani and G. C. Samanta {\it Int. Jour. Mod. Phys. D,} {\bf 28}, 1950039 (2019)

\bibitem{ann} P. Sahoo, A. Kirschner and P.K. Sahoo, {\it Mod. Phys. Lett. A,} {\bf 34}, 1950303 (2019).

\bibitem{moraes} P.H.R.S. Moraes, W. de Paula, R.A.C. Correa, {\it Int. J. Mod. Phys. D.} {\bf 30}, 13 (2021).

\bibitem{mandal} P. Sahoo, S. Mandal and P. K. Sahoo, {\it New Astronomy.} {\bf 80}, 101421 (2020)

\bibitem{galin} G. Gyulchev, P. Nedkova, V. Tinchev and S. Yazadjiev, {\it Eur. Phys. J. C.} {\bf 78}, 544 (2018)

\bibitem{rah} F. Rahaman, P. K. F. Kuhfittig, S. Ray and N. Islam, {\it Eur. Phys.
J. C.} {\bf 74}, 2750 (2014).

\bibitem{tang} Z. Xu, M. Tang, G. Cao and S. Zhang, {\it Eur. Phys.
J. C.} {\bf 80}, 70 (2020).

\bibitem{eu} L. P. Euvé and G. Rousseaux  {\it Phys. Rev D.} {\bf 96}, 064042 (2017).

\bibitem{rom} G.E. Romero, R. Thomas and D. Pérez, {\it Int. J. Theor. Phys.} {\bf 51}, 925-942 (2012)


\bibitem{termworm} M. Rehman and K. Saifullah, {\it J. Cosmol. Astropart. Phys.} {\bf 020}, JCAP06 (2021)


\bibitem{reiss} A.G.Reiss, et al. {\it Astrophys. J.} {\bf 607}, 665 (2004).

\bibitem{hong} S. Hong and S. Kim, {\it Mod. Phys. Lett. A.} {\bf 21}, 10 (2006)


\bibitem{dai} D. Dai and D. Stojkovic, {\it Phys. Rev. D.} {\bf 100}, 083513 (2019)

\bibitem{lobo} F. S. N. Lobo, arXiv:0710.4474v1.

\bibitem{ford} L. H. Ford and T. A. Roman {\it Phys. Rev D.} {\bf 51}, 8 (1995).

\bibitem{dad} M. Visser, S. Kar and N. Dadhich {\it Phys. Rev. Lett.} {\bf 90}, 20 (2003).

\bibitem{lobo3} F.S.N. Lobo, {\it Phys. Rev. D.} {\bf 71}, 084011 (2005)


\bibitem{dad} M. Visser, S. Kar and N. Dadhich {\it Phys. Rev. Lett.} {\bf 90}, 20 (2003).

\bibitem{cle2} L. Chetouani, G. Clément {\it General Relativity and Gravitation.} {\bf 16}, 2 (1984)


\bibitem{an} A. Das, A. Saha and S. Gangopadhyay, {\it J. Cosmol. Astropart. Phys.} {\bf 06} (2021)


\bibitem{reg} R. C. Pantig, P. K. Yu, E. T. Rodulfo and A. Övgün, {\it Ann. Phys.} {\bf 436} (2022)

\bibitem{javed} W. Javed, A. Hamza and A. Övgün, {\it Mod. Phys. Lett. A} {\bf 35}, 39 (2020).

\bibitem{kuf} P. Kuhfittig, {\it Cent. Eur. J. Phys.} {\bf 9}, 5 (2011)

\bibitem{godsam} N. Godani and G. C. Samanta, {\it  Int. Jour. G. M. in Mod. Phys.} {\bf 18}, 7 (2021)


\bibitem{sah} P. Sahoo, P. H. R. S. Moraes, M. M. Lapola and P. K. Sahoo, {\it Int. J. Mod. Phys. D.} {\bf 28}, 8 (2019)

\bibitem{kimet} K. Jusufi, A. Ovgün, A.Banerjee, {\it Eur. Phys. J. Plus} {\bf 134}, 428 (2019)

\bibitem{synge} J. L. Synge, {\it Mon. Not. Roy. Astron. Soc.} {\bf 131}, 3, 463 (1966).

\bibitem{ferrari} V. Ferrari, L. Gualtieri, P. Pano {\it General Relativity
and its Applications}, 1st edn. (CRC press, 2021).

\bibitem{carter} B. Carter, {\it Phys. Rev.} {\bf 174}, 1559 (1968).

\bibitem{chandra} S. Chandrasekhar, {\it The Mathematical Theory of Black Holes}, 1st edn. (Oxford University Press, 1983).

\bibitem{pelle2022skylight} J. Pelle, O. Reula, F. Carrasco, C. Bederian, {\it Monthly Notices of the Royal Astronomical Society} {\bf 515}(1), 1316-1327 (2022)

\bibitem{raj+19} S. Rajibul, B. Pritam, P. Suvankar and S. Tapobrata, {\it Journal of Cosmology and Astroparticle Physics} {\bf 7}, 028 (2019)

\bibitem{sak+14}N. Sakai, H. Saida, T. Tamaki, {\it Phys. Rev. D} {\bf 90}, 104013 (2014)

\bibitem{kub+16}T. Kubo, N. Sakai, {\it Phys. Rev. D} {\bf 93}, 084051 (2016)

\bibitem{vis+04}M. Visser and D. L. Wiltshire, {\it Classical Quantum Gravity}
{\bf 21}, 1135 (2004).



\end{thebibliography}


\end{document}